\documentclass{article}

     \PassOptionsToPackage{numbers, compress}{natbib}

     \usepackage[final]{neurips_2024}

\usepackage[utf8]{inputenc} 
\usepackage[T1]{fontenc}    
\usepackage{hyperref}       
\usepackage{url}            
\usepackage{booktabs}       
\usepackage{amsfonts}       
\usepackage{nicefrac}       
\usepackage{microtype}      
\usepackage{xcolor}         
\usepackage{amsmath}        
\usepackage{graphicx}
\usepackage{tabularx}

\title{Fine-Tuning Large Language Models for Educational Support: Leveraging Gagné's Nine Events of Instruction for Lesson Planning
\thanks{Jia, L., Qi, C., Wei, Y., Sun, H, Yang, X. (2024). Fine-Tuning Large Language Models for Educational Support: Leveraging Gagné's Nine Events of Instruction for Lesson Planning. Conference Proceedings of the 28th Global Chinese Conference on Computers in Education (GCCCE 2024), 62–69. Chongqing, China: Global Chinese Conference on Computers in Education.}}

\author{
Linzhao Jia$^{1, 2, 3, 4}$, Changyong Qi$^{1, 2, 3}$, Yuang Wei$^{1, 2, 3}$, Han Sun $^{1, 2, 3, 4}$ , Xiaozhe Yang $^{1, 2, 4}$\thanks{Corresponding Author: lzjia@stu.ecnu.edu.cn} \,\,\\
\\
$^{1}$~\text{Lab of Artificial Intelligence for Education, East China Normal University}\\
$^{2}$~\text{Shanghai Institute of Artificial Intelligence for Education, East China Normal University}\\
$^{3}$~\text{School of Computer Science and Technology, East China Normal University}\\
$^{4}$~\text{Institute of Curriculum and Instruction \& Classroom Analysis Lab, East China Normal University}\\
}
\begin{document}
\maketitle
\begin{abstract}
Effective lesson planning is crucial in education process, serving as the cornerstone for high-quality teaching and the cultivation of a conducive learning atmosphere. This study investigates how large language models (LLMs) can enhance teacher preparation by incorporating them with Gagné’s Nine Events of Instruction, especially in the field of mathematics education in compulsory education. It investigates two distinct methodologies: the development of Chain of Thought (CoT) prompts to direct LLMs in generating content that aligns with instructional events, and the application of fine-tuning approaches like Low-Rank Adaptation (LoRA) to enhance model performance. This research starts with creating a comprehensive dataset based on math curriculum standards and Gagné's instructional events. The first method involves crafting CoT-optimized prompts to generate detailed, logically coherent responses from LLMs, improving their ability to create educationally relevant content. The second method uses specialized datasets to fine-tune open-source models, enhancing their educational content generation and analysis capabilities. This study contributes to the evolving dialogue on the integration of AI in education, illustrating innovative strategies for leveraging LLMs to bolster teaching and learning processes.
\end{abstract}

\section{Introduction}
The advent of large language models (LLMs) has revolutionized various domains, notably in education \cite{li_generating_2024, jiang_ai_2024}. This development is not only reshaping educational methodologies and analytics but also highlights a growing trend: an increasing number of individuals are leveraging LLMs \cite{forero2023chatgpt, sun2023chatgpt, uchiyama2023large}, including tools like ChatGPT, to analyze teacher dialogue in classroom settings \cite{hu_finetuning_2024, liu_educational_2024} or assist teachers in lesson preparation,\cite{zhou_study_2024,jiang_multi-agent_2024}. However, the current landscape lacks a streamlined pipeline for educators to efficiently leverage LLMs, posing challenges in directly applying these models for instructional support. Given that LLMs are not specifically trained on educational datasets, educators are required to invest time in designing prompts, which may not always yield satisfactory results. Our research aims to bridge this gap by proposing a comprehensive pipeline that not only standardizes the use of LLMs in educational settings but also optimizes them for educational content, enhancing their utility in teaching design and evaluation. 
In this study, we aim to fine-tune LLMs through prompt tuning and model tuning within the framework of Gagné's Nine Events of Instruction, thereby assisting teachers in classroom dialogue assessment or in generating classroom dialogue teaching plans. Robert Gagné is considered to be the foremost researcher and contributor to the systematic approach to instructional design and training \cite{kruse2009gagne}. By integrating Gagné's Nine Events of instruction into the design of classroom dialogue, educators can leverage a proven instructional design framework to enhance the effectiveness of their teaching, aligning closely with the cognitive, affective, and psychomotor objectives outlined in Bloom's Taxonomy.
For prompt tuning, we design scenarios that prompt non-open-source models, like ChatGPT \cite{achiam2023gpt}, to guide the LLMs to generate dialogue that enriches teaching and learning experiences. As shown in Figure \ref{fig1}, through a comparative analysis of CoT and direct prompts, this research aims to illuminate the benefits of employing structured, thought-provoking prompts to optimize the educational utility of language models in creating and evaluating teaching dialogue. As for model tuning, we fine-tuned on open-source models, including ChatGLM \cite{du2021glm} and Llama-Chinese \cite{touvron2023llama}. Training these LLMs on our specially designed dataset imbues them with a deeper understanding of instructional design, significantly enhancing their capacity to produce content that is not only relevant but also robust. Our contributions to the field of instructional technology and educational methodology enhancement can be summarized as follows:
\begin{figure}[t]
    \centering
    \includegraphics[width=\linewidth]{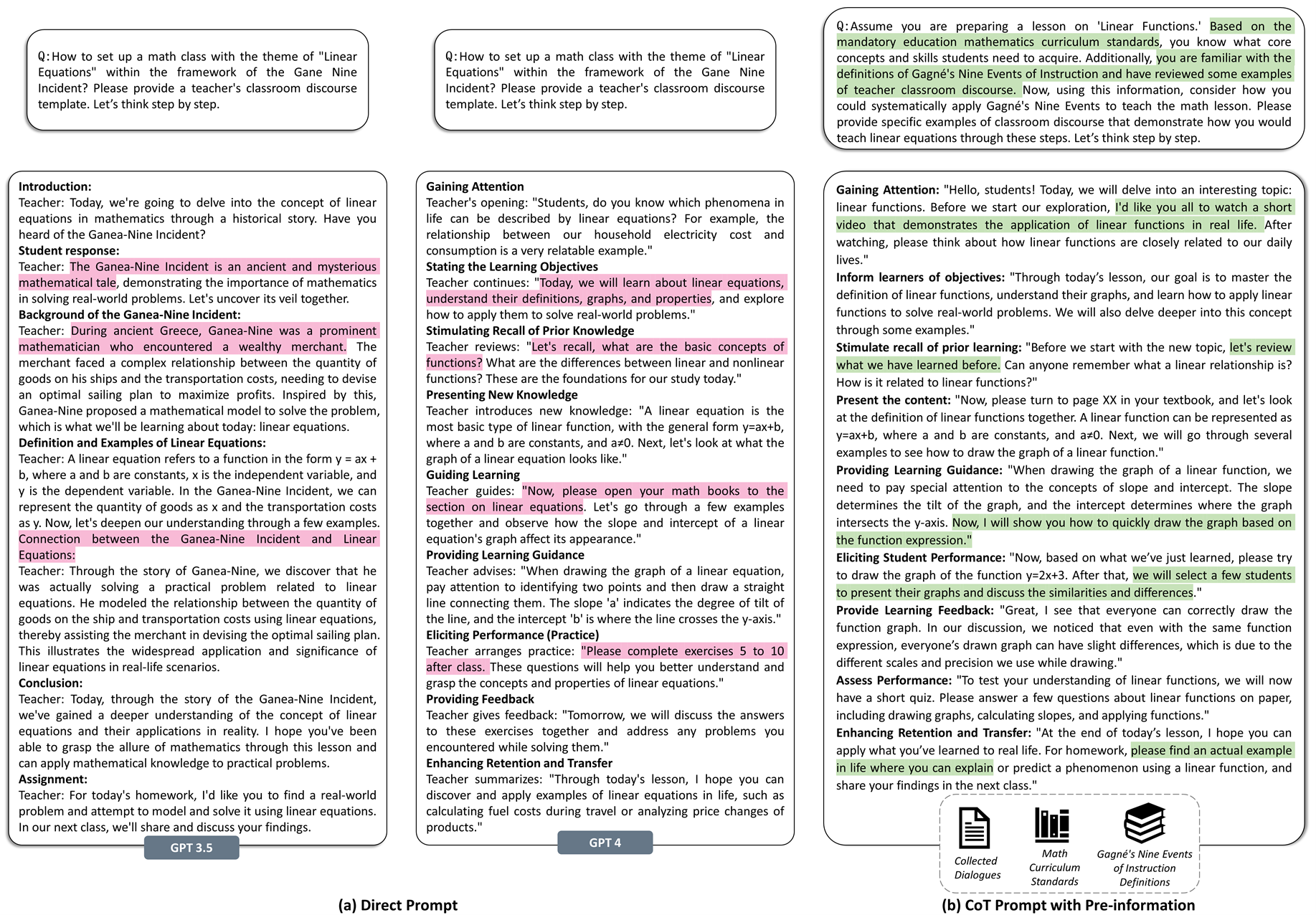}
    \caption{Comparative analysis of CoT and direct prompts.}
    \label{fig1}
\end{figure}
\begin{enumerate}
    \item \textbf{Open-access dataset:} We have developed an open-access dataset specifically tailored for training and evaluating LLMs within the context of educational dialogue generation. This dataset is designed to encapsulate a wide range of instructional scenarios, facilitating the development of LLMs capable of understanding and generating pedagogically valuable content.
    \item \textbf{Novelty Pipeline:} Our research introduces a novel pipeline for employing LLMs as central tools in advancing educational methodologies. By integrating structured prompt tuning and fine-tuning processes, we aim to unlock the full potential of LLMs in creating dynamic, interactive, and enriching teaching and learning experiences.
    \item \textbf{Fine-tuned LLM:} The fine-tuned LLMs resulting from our research can be seamlessly integrated into classroom analysis systems. These models have been specifically enhanced to comprehend and generate educational content, making them invaluable assets for teachers and educational content creators. By embedding these models within classroom analysis systems, educators can leverage real-time, AI-driven insights to tailor teaching strategies and content, thereby optimizing the teaching impact.
\end{enumerate}

\section{Related Work}

\subsection{Large Language Models in Education}
The integration of LLMs into educational practices represents a significant leap forward in the use of artificial intelligence to enhance teaching and learning processes. The study \cite{golchin2024large} assesses GPT-4 and GPT-3.5's grading of MOOC courses using enhanced Zero-shot-CoT prompting, showing GPT-4's superior alignment with instructor grades, particularly in subjects requiring imaginative thinking. Long et al., explored the application of GPT-4 in labeling classroom dialogues \cite{long2024evaluating}. The findings demonstrate that employing GPT-4 for this task not only achieves a high degree of similarity with human labeling but also significantly reduces the time required for the process. Qian et al., \cite{qian2023user} developed a user-adaptive generative chatbot that facilitates language learning through constrained decoding, dynamically aligning its curriculum to user requests. Macina introduces MATHDIAL, a semi-synthetic dialogue dataset for math tutoring that enhances pedagogical effectiveness through equitable practices and scaffolding, offering over 3,000 conversations for benchmarking and advancing tutor response generation models, and setting a new standard in dialogue-based tutoring research in NLP \cite{macina2023mathdial}.

\subsection{Prompt Tuning and Model Tuning in Educational Settings}

Recent studies have illustrated that augmenting the number of tasks within constructed instructions can enhance the generalization capabilities of LLMs. Notably, prompting with Chain-of-Thought (CoT) \cite{wei2022chain} has been proven to effectively tackle intricate problems by systematically producing steps of reasoning. This methodology's success has sparked significant interest in the educational sector. Specifically, the Socratic teaching method \cite{dan2023educhat}, which guides students towards self-discovery and understanding by asking progressively complex questions, mirroring the stepwise elucidation process found in CoT prompting. Building on these insights, a study \cite{uchiyama2023large} presents a system leveraging LLMs, notably ChatGPT, to facilitate preparation learning in flipped classrooms, featuring automatic student query responses and enhanced prompt clarity through video subtitles, aiming to mitigate common flipped classroom challenges and contribute to learning analytics through data collection. 
In the domain of educational technology, the fine-tuning of large models has gained significant attention. Among these advancements, Dan et al. have introduced EduChat, notable for its innovative approach to interactive learning using fine-tuned language models. This system is distinguished by its ability to tailor educational content and provide dynamic, context-aware interactions with students. Collectively, these studies embody diverse strategies for leveraging language models, each offering unique perspectives on the depth, relevance, and educational utility of the generated content, thereby enriching the landscape of LLM application in educational technology.

\section{Framework}

In this section, we discuss the dataset preparation and fine-tuning processes, which are essential for customizing LLMs to adeptly analyze and generate teacher dialogue aligned with Gagné's Nine Events of Instruction.

\subsection{Dataset Preparation}
The initial input data we used in this study mainly consists of three parts, (1) Definitions and examples of Gagné's Nine Events of Instruction, which serving as a foundational guide for model training and evaluation. (2) Math Curriculum Standards: By obtaining the official published curriculum standards and using manual extraction methods, the curriculum requirements for each knowledge point are organized and input to LLMs as a limiting condition for generating teacher dialogue templates. (3) Collected Dialogues: Extract teacher dialogues and teacher-student interaction from real classroom environments, and manually annotate them as examples of real-life Gagné's teaching events. To obtain more diversity and comprehensiveness data, we manually write prompt templates and have GPT4 generate some examples of teacher dialogue templates. These prompts are designed to guide GPT-4 in generating teacher dialogue that is not only subject-specific but also structured around Gagné's instructional framework.

\begin{figure}[t]
    \centering
    \includegraphics[width=\linewidth]{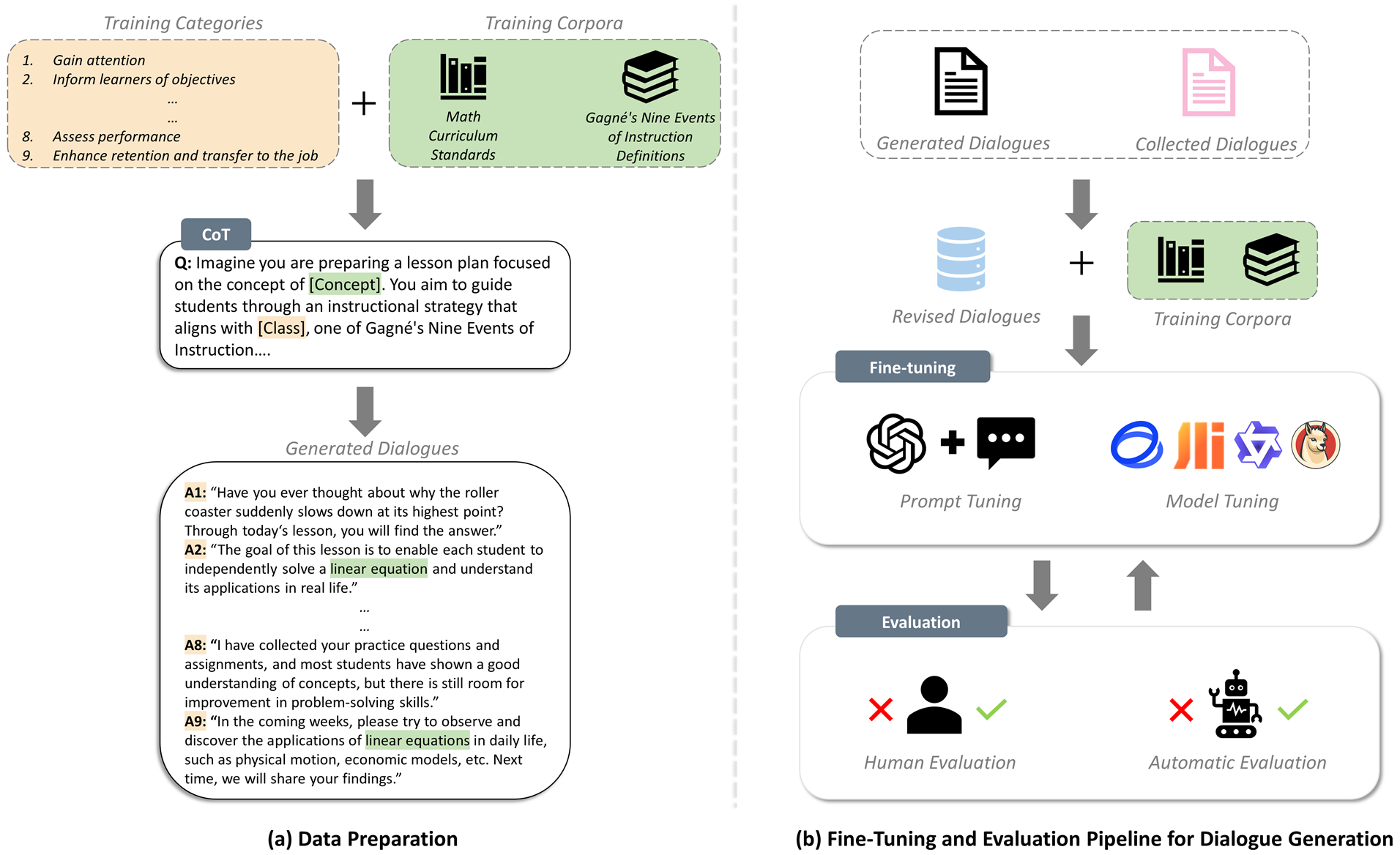}
    \caption{Illustration of the process for generating educational dialogue via LLMs.}
    \label{fig2}
\end{figure}

As shown in Figure \ref{fig2}.(a), the CoT prompt template is designed to lead the language model through a comprehensive thought process that covers all of Gagné's Nine Events, tailored to teaching the mathematical concept of linear equations. By replacing [Concept] with " linear equation " and adapting the template to each specific event [Class], the model can generate detailed and instructionally sound teacher dialogue for each stage of the lesson.
After the data generation phase, a team of educators and subject matter experts conducts a thorough review of GPT-4's outputs. This involves checking for accuracy, alignment with the curriculum standards, and adherence to Gagné's Nine Events. Any inaccuracies or misalignments are manually corrected or removed to ensure the final dataset comprises high-quality teacher dialogue templates. In Table 1, we present examples of the speech templates generated based on Gagné's nine events, along with a statistical count of each event. Due to the dataset being manually curated and revised, the quantities across categories are not uniform.

\begin{table}[t]
    \centering
    \caption{Generated dataset.}
    \begin{tabularx}{\textwidth}{lXl}
    \toprule
        Gagné's Nine Events of Instruction & Example & Number\\
        \hline
       Gain attention  & Let's begin with a quick challenge: Can anyone guess which shape we’ll be exploring today with this image? & 715 \\
       Inform learners of objectives & Today, we'll learn how to calculate the area of different triangles, a skill essential for many real-life applications. & 422 \\
       Stimulate recall of prior learning & Think back to last week when we studied polygons. How does understanding their properties help in calculating the area of triangles? & 769 \\
       Present the content & We’ll look at formulas for the area of right, isosceles, and equilateral triangles using visuals and demonstrations. & 768 \\
       Provide ''learning guidance'' & I’ll guide you through each formula step-by-step. Remember, the base and height are key in these calculations. & 862 \\
       Elicit performance & Now, let's try calculating the area of these triangles together. Start with this right triangle. & 680 \\
       Provide feedback & That’s a good attempt. Make sure to measure the height correctly—it’s perpendicular to the base. & 823 \\
       Assess performance & I’ll hand out a short quiz now. You'll calculate the area of a few triangles to show your understanding. & 476 \\
       Enhance retention and transfer to the job & As homework, find objects at home shaped like triangles and calculate their area. It’ll help solidify what we learned today. & 665 \\
       \bottomrule
    \end{tabularx}
    \label{tab:my_label}
\end{table}

\subsection{Fine-tuning}
\subsubsection{Prompt tuning}
To enable LLMs to better handle teacher dialogue generation tasks that comply with the Gagné's Nine Events, we first build the agent by utilizing the ChatGPT API and connecting downstream tasks to complete this task. As shown in Figure \ref{fig2}.(b), we take the revised dialogues and training corpus as input, and employ the CoT method to guide the ChatGPT model towards generating outputs that adhere to Gagné's Nine Events. The CoT approach involves structuring prompts that not only direct the model's attention to the task but also encourage it to follow a logical sequence of thoughts. These prompts are designed to elicit detailed and structured responses from ChatGPT, ensuring that the generated dialogue is both pedagogically sound and closely aligned with the instructional objectives outlined in the curriculum standards.
For instance, to generate content that aligns with the 'Gain Attention' event, the CoT prompt might begin with a real-world problem that can be solved using the concept being taught, followed by questions or scenarios that lead ChatGPT to explore and elucidate the concept in a manner that would captivate students’ interest. The output from ChatGPT, post-CoT prompting, is then evaluated for alignment with the instructional principles of Gagné's Nine Events. Through iterative refinement and tuning of the CoT prompts, we optimize the model's performance, ensuring that the generated teacher dialogues are aligned with the desired instructional outcomes. This methodological approach demonstrates a novel application of LLMs in educational content creation, enhancing the quality and efficacy of teacher-student interactions in the learning process.

\subsubsection{Model tuning}
In the realm of model fine-tuning for enhancing the generation of teacher classroom dialogue aligned with Gagné's Nine Events of Instruction, we adopt a nuanced approach that diverges from traditional extensive corpus retraining methods. Our strategy involves selectively adjusting critical parameters within the ChatGLM model, leveraging a retrieved dataset meticulously constructed to embody the essence of educational dialogue as delineated by Gagné's framework. Initially, we identify the pivotal aspects of Gagné's Nine Events that are most crucial for effective classroom dialogue. The constructed dataset serves as a comprehensive resource, embedding these instructional principles within authentic and synthesized educational contexts. This dataset not only provides the model with a rich variety of pedagogically aligned content but also exposes it to the nuanced dynamics of classroom interaction.
Subsequently, employing the ChatGLM model as our foundational framework, we delve into fine-tuning, specifically employing Low-Rank Adaptation (LoRA) techniques. This approach enables us to refine the model's parameters subtly, enhancing its ability to process and produce content that resonates with the instructional strategies inherent in Gagné's Nine Events. The ultimate goal is to validate the model's enhanced capability through a structured experimental workflow, assessing its performance across various educational dialogues and scenarios. This fine-tuning endeavor marks a significant stride towards realizing the potential of LLMs in education, promising a future where AI-driven tools synergize with pedagogical methodologies to foster enriched learning experiences.

\section{Experiments}
\subsection{Dataset}
The dataset, central to our study, is meticulously crafted around Gagné's Nine Events of Instruction and includes a rich blend of content generated by ChatGPT and real teacher utterances gathered from live classroom settings. This comprehensive collection forms the backbone of our research, offering a nuanced perspective on educational dialogue. These utterances have been meticulously annotated and refined manually, providing a rich and diverse basis for our analysis.

\subsection{Baseline Models}
In the process of developing our prompt tuning model, we utilized GPT3.5 and GPT-4 \cite{brown2020language, achiam2023gpt}. GPT-3.5, released by OpenAI, is acclaimed for its exceptional generative power and wide-ranging applications, symbolizing a crucial advancement in language model development. Following this, GPT-4 was introduced with substantial refinements in model architecture, training dataset breadth, and diversity, thereby significantly boosting the performance capabilities of the model.

Continuing, we turn our attention to open-access models that can be deployed locally, including Llama2-7B \cite{touvron2023llama}, ChatGLM3-7B \cite{du2021glm}, Qwen-7B \cite{bai2023qwen}, and Baichuan2-7B \cite{yang2023baichuan}. These models are not only innovative in their technical design but also offer the flexibility of local deployment due to their open-access nature, facilitating research and practical applications. We conducted fine-tuning on these models using the Llama-Factory \cite{zheng2024llamafactory} platform.

\subsection{Evaluation Metric}
Evaluate the models' efficacy through quantitative metrics and undertake human assessments by involving frontline educators to appraise the models' outputs for their consistency with Gagné's events and the overall quality of instruction.
\begin{itemize}
    \item \textbf{BLEU-4 (Bilingual Evaluation Understudy):} BLEU-4 is used to evaluate the quality of machine-translated text against a set of reference translations. It focuses on the accuracy of four consecutive words (4-grams) in the translated text. A higher BLEU score indicates better translation quality.
    \item \textbf{ROUGE (Recall-Oriented Understudy for Gisting Evaluation):} ROUGE is a set of metrics for evaluating automatic text summarization and machine translation by comparing the overlap between the generated text and reference texts. This set encompasses several variants: ROUGE-1, which quantifies the overlap of unigrams, thus assessing the extent of basic content similarity; ROUGE-2, which measures the overlap of bigrams, offering insights into the prevalence of contiguous word pairs; and ROUGE-L, which evaluates the longest common subsequence, serving as an indicator of the text's coherence over extended sequences.
\end{itemize}

\subsection{Human Evaluation}
To validate the practical application of our method, we are designing a human evaluation study where frontline teachers will assess the quality and applicability of the LLM-generated content in real-world classroom settings. We intend to create a questionnaire focusing on the following key dimensions:
\begin{itemize}
    \item \textbf{Alignment with Instructional Principles:} We will examine the extent to which the generated dialogue adheres to  Gagné's Nine Events of Instruction, evaluating its consistency with established educational methodologies.
\item \textbf{Content Quality:} The pedagogical soundness, accuracy, and alignment with the math curriculum standards will be scrutinized to ensure the content's educational validity.
\item \textbf{Engagement and Clarity:} This criterion focuses on the dialogue's ability to capture students' interest and clearly communicate mathematical concepts, pivotal for fostering an effective learning environment.

\end{itemize}
By meticulously examining these key dimensions, we aim to bridge the gap between advanced language technology and real-world educational needs, ensuring that the innovative potential of LLMs is fully realized and effectively harnessed to enhance teaching and learning outcomes.

\section{Discussion}
This study aims to verify whether LLMs can effectively analyze and generate teacher classroom dialogue that aligns with Gagné's Nine Events through precise prompt design or dataset fine-tuning. The model can be instrumental in curriculum development and assessment design, ensuring that educational objectives are met effectively across different cognitive domains. And this blend not only provides a more nuanced understanding of pedagogical effectiveness but also offers a scalable method to enrich teaching materials.
The findings from our study illuminate the substantial potential of LLMs like ChatGPT in enriching educational content creation. Moving forward, it is imperative to continue refining these models, ensuring they are both pedagogically sound and practically applicable. Collaborations between AI researchers and educational practitioners will be key in achieving a symbiotic relationship between technology and teaching, ultimately leading to enhanced educational outcomes.

\section{Conclusion}
In summary, our research endeavors to bridge the gap between the capabilities of current LLMs and the evolving needs of educational practitioners. By providing open-access resources, introducing a novel application pipeline, and delivering fine-tuned models ready for integration into educational systems, we contribute substantially to the enhancement of instructional technologies and methodologies. Our exploration into the use of LLMs for generating and analyzing teacher dialogue within the framework of Gagné's Nine Events of Instruction has underscored the viability and effectiveness of these models in educational settings. The successful application of both prompt tuning and model fine-tuning strategies has set the groundwork for future research, particularly in the systematic integration of AI in curriculum design and instructional evaluation.

\bibliographystyle{IEEEtran}
\bibliography{sample-base}

\end{document}